\documentclass[11pt,letter]{article}

\usepackage{poma_style}
\usepackage[utf8]{inputenc}
\usepackage[T1]{fontenc}
\usepackage{amsmath}
\usepackage{graphicx}
\usepackage{subcaption}
\PassOptionsToPackage{hyphens}{url}
\usepackage[colorlinks=true,urlcolor=blue,linkcolor=black,citecolor=black]{hyperref}
\usepackage{float}

\newif\ifanonymous
\anonymousfalse   

\usepackage{fancyhdr}
\renewcommand{\headrulewidth}{0.4pt}

\usepackage{tikz}

\title{Digital Revival: Acoustic Documentation and Digital Reactivation of Historical Woodwind Instruments}

\begin{document}

\maketitle
\thispagestyle{fancy}

\begin{figure}[H]
  \centering
  \begin{subfigure}{\columnwidth}
    \centering
    \includegraphics[width=\textwidth]{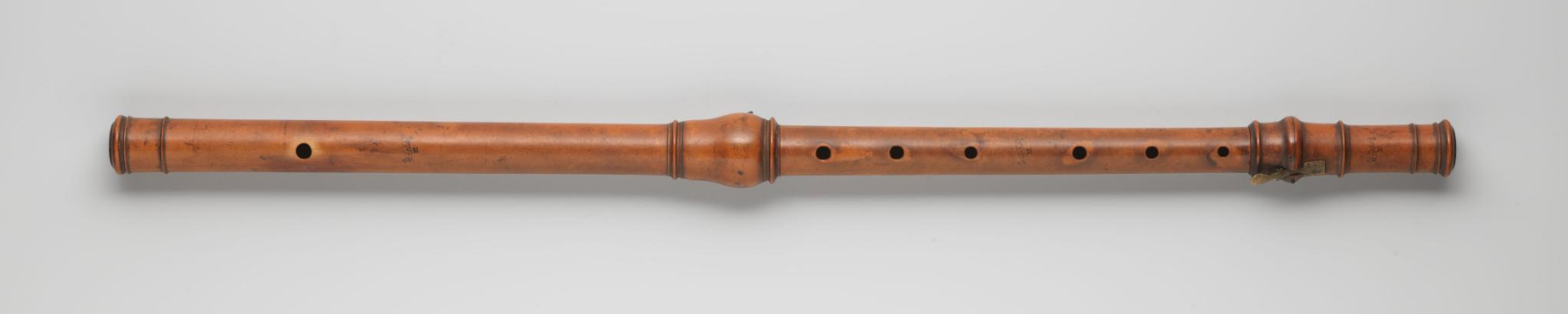}
    \caption{The Haka flute}
    \label{fig:haka}
  \end{subfigure}
  \vspace{4pt}
  \begin{subfigure}{\columnwidth}
    \centering
    \includegraphics[width=\textwidth]{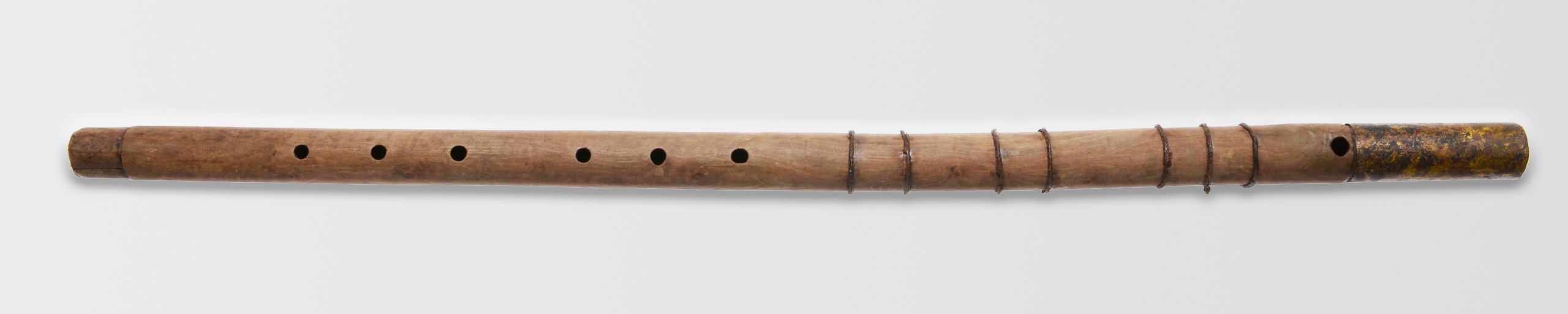}
    \caption{The Warder flute}
    \label{fig:warder}
  \end{subfigure}
  \caption{The two instruments studied in this paper. (a)~The Haka flute, c.~1680, attributed to Richard Haka, CollectieCentrum Nederland: a transverse flute representing a transitional point between Renaissance and Baroque traverso traditions (photo: Rijksmuseum, Amsterdam). (b)~The Warder flute, c.~1540, builder unknown, Huis van Hilde museum, Castricum, the Netherlands: recovered from a shipwreck in North Holland in 2017, considered one of the oldest surviving flutes found in the Netherlands.}
  \label{fig:flutes}
\end{figure}

\section{Introduction}

Historical musical instruments are objects of fascination, encompassing in their identity craft, culture, heritage and sound. Many instruments held in museum collections remain unplayable due to age, fragility, or conservation restrictions, leaving their sound inaccessible to viewers, researchers and performers alike. The Digital Revival project (stylized as \textit{dgtalrevival})~\ifanonymous\cite{anon2025digitalrevival}\else\cite{digitalrevival2025}\fi\ addresses this gap by capturing the sounds of historical European woodwind instruments and reactivating them as playable digital instruments, performed via an Electronic Wind Instrument (EWI) and released as musical recordings.

\subsection{Historical Instruments and Acoustic Heritage}

The sonic dimension of cultural heritage has gained increasing recognition as a research domain in its own right. Katz et al.~\cite{katz2024pasthasears} 
developed acoustic simulations of the Cath\'edrale Notre-Dame de Paris. Using geometric data, they reconstructed the cathedral's acoustic environment at different historical stages, each corresponding to major architectural changes to the structure. The simulated acoustic environments were used to produce pre-rendered binaural audio content intended for the general public, distributed as a podcast series and a mobile audio guide.

The NEMUS project represents the most systematic effort to date in instrument revival, combining numerical acoustic methods and physical modelling to restore the sound of historical instruments. The project uses physical measurements and geometric data, and produces audible acoustic reconstructions of museum specimens, including Renaissance recorders, Baroque oboes, and early clarinets~\cite{nemus2021project}. 

\subsection{Non-Invasive Characterization and Replica Construction}

Historical instruments that cannot be played can still be acoustically characterized through non-invasive measurement techniques. Eveno and Le~Conte measured the input impedance of 45 serpents from the collection of the Mus\'{e}e de la Musique, spanning instruments of different shapes, numbers of holes, and makers~\cite{eveno2016acoustical}. 
Measurements were performed using a piezoelectric buzzer as the acoustic source; two microphones measured pressure and acoustic flow at the instrument's input, and their transfer function yielded the impedance. Consistent resonance characteristics were found across specimens, validating input impedance measurement as a reliable non-invasive method. This method is applicable to instruments in playable condition. Bowen et al.\ validated a complementary approach: bore geometry measurements were used to compute the acoustic impedance of a historical bass clarinet, and the predictions were cross-checked against measured impedance and playing tests on the same instrument, which had been kept in playable condition~\cite{bowen2019assessing}. 
This establishes that bore geometry alone may be sufficient to reliably infer pitch, intonation, and timbral characteristics, even for instruments in non-playable condition.

For instruments where playing is not permitted or possible, the primary characterization method is geometric imaging. X-ray computed tomography has been applied to historical woodwind instruments, including a piccolo and a baroque transverse flute from a private collection, to extract bore geometry, wall thickness, and manufacturing traces without disassembly~\cite{tansella2022x}. 
The resulting geometric data was used to produce acoustically faithful 3D-printed replicas.

Geometric scan data can be used to fabricate physical replicas and to inform physical modelling. Fritz et al.\ produced 3D-printed copies of a Hotteterre traverso flute from the Mus\'ee de la Musique using stereolithography, with geometry derived from X-ray tomography, and evaluated perceptual differences against a wooden facsimile~\cite{fritz2025museum}. 
Players perceived the wooden facsimile as richer, warmer, and more authentic, while the 3D-printed copy was described as overly easy and homogeneous; however, in listening tests with 69 participants, neither experts nor non-experts could reliably distinguish the two instruments by ear, with discrimination performance near chance level.
Further work has demonstrated 3D printing as a viable fabrication method for other wind instruments, including cornetts and flutes~\cite{savan2014cad,zoran2011flute}. 
Beyond replication, digital modelling of historical instruments can reveal the manufacturing tools used in their making: Waters reconstructed the geometry of a tonehole undercutting tool (fraise) from the tonehole shapes of Richard Potter flutes, arguing that a complete understanding of a historical instrument requires conceiving not only the object but the tools that produced it~\cite{waters2025modeling}. 

\subsection{Physical Modelling of Historical Woodwind Instruments}

The existing understanding of flue instrument acoustics permits a convincing synthesis of the instrument sounds by physical modelling.
The acoustic behavior of a transverse flute is governed principally by its bore geometry: the length and cross-sectional profile of the air column, together with tonehole positions and dimensions, determine the resonance structure and thus intonation, range, and timbre~\cite{fletcher1998physics}. 

Wall material plays a secondary role: while viscous and thermal losses at the bore surface vary with material, their effect on tone is minor. Thus, tonal differences between instruments of different materials are largely attributable to geometric variation rather than the material itself~\cite{fletcher1998physics}. 
Physical models derived from geometric data are therefore informative even when material parameters are uncertain or have changed due to aging. Aeroacoustic excitation in flue instruments, including jet formation, edge-tone coupling, and the role of turbulent flow in producing characteristic onset transients, has been treated comprehensively~\cite{fabre2012aeroacoustics}. 

For bores that deviate substantially from uniform cylindrical or conical profiles (including Renaissance traversos with gradually varying tapers and deformed cross-sections), multi-modal waveguide theory extends the standard plane-wave transfer matrix. Multi-modal acoustic propagation in varying and non-circular cross-section waveguides has been rigorously formulated and validated~\cite{pagneux1996waveguide1,amir1997waveguide2}, 
and is directly applicable to the Warder flute, whose bore has been deformed by waterlogged burial.

Systematic acoustic differences between baroque, classical, and modern flutes, attributable to bore profile, tonehole geometry, and embouchure design, have been characterized~\cite{wolfe2001impedance}, 
providing a comparative reference framework applicable to the instruments discussed in this paper.

\subsection{Acoustic Aspects of Wood Aging and Waterlogging}

The material properties of wood are not static. Natural aging reduces the wood's sensitivity to ambient moisture while improving its acoustic quality, with measurable increases in sound velocity and decreases in internal friction, attributable to cellulose recrystallization and hemicellulose depolymerization~\cite{obataya2017ageing,gurau2023aging}. 
In instruments where the wooden components vibrate and act as resonators or radiators, such as many string instruments, material changes have a direct and audible effect on tone. In flutes, however, the acoustic behavior is dominated by the air column, and wall material plays a secondary role; these aging-induced changes in material properties therefore may have a measurable but minor acoustic consequence.

Waterlogged burial is the long-term immersion in water-saturated sediment under anaerobic conditions, where the absence of oxygen prevents complete biological decay. This condition imposes more severe material change. Slow microbial degradation of cell walls and progressive loss of structural polymers alter the wood's mechanical state over centuries~\cite{bjordal2021waterlogged}. 
Post-excavation drying disrupts the moisture equilibrium established during burial, resulting in uneven shrinkage, cracking, and bore deformation~\cite{bjordal2021waterlogged}. 

\subsection{The Digital Revival Project}

The Digital Revival project transforms historical European woodwind instruments, many of them unplayable in their museum state, into expressive digital entities performed via modern wind controllers. Rather than pursuing exact acoustic replication, its guiding philosophy is the capture of the \textit{sonic DNA} of historical sources, enabling new musical identities to emerge from them. The methodology combines high-resolution sampling, performance modelling, and real-time EWI control; for instruments that cannot be played, physical and spectral analysis take on a more central role.

Instruments addressed so far include the Richard Haka flute and oboe, a Claude Laurent crystal flute, a series of Engelbert Terton recorders, and the Warder flute, drawn from the collections of the Rijksmuseum, the Kunstmuseum Den Haag, and Huis van Hilde. Outputs include live performances, recordings, and hybrid works combining sampling with physical modelling. The project is accompanied by a public-facing website\ifanonymous\ [anonymized for review]~\cite{anon2025instrumenta}\else, Instrumenta Online~\cite{instrumenta2025},\fi\ that makes revived instruments available for online performance and documents the project's ongoing work.

\section{Digital Revival of a Functional Flute: Haka Case Study}

This case study demonstrates the application of the Digital Revival approach to a functional historical instrument, termed the \textit{Haka flute}, illustrating the practical constraints and design decisions characteristic of real-world revival projects.

The Haka flute, shown in Fig.~\ref{fig:flutes}(a), is a transverse flute, dated to approximately 1680, made by Am\-ster\-dam-based instrument maker Richard Haka (1646--1705)~\cite{vanacht1988dutch}. It is held in the collection of the CollectieCentrum Nederland (Collection Centre of the Netherlands), and occasionally on display at the Rijksmuseum, Amsterdam. Haka occupies a central position in the early Dutch development of the transverse flute; instruments from his workshop represent a transitional point between Renaissance and Baroque traverso traditions. A Renaissance traverso has a cylindrical bore, while a Baroque traverso is conical, tapering towards the foot. The Haka flute lies between these: both the head joint and the long middle section are predominantly cylindrical, tapering only very slightly toward their lower ends, completed by a separate foot joint. A modern replica in boxwood was constructed by instrument builder Simon Polak~\cite{polak2025earlyflute}. The Haka flute has remained sufficiently intact to permit limited controlled playing. Thus, it is a well-suited candidate for this project.

\subsection{Recording Procedure}

Recording was conducted under significant technical and conservation constraints. The instrument was not permitted to be transported to a dedicated studio; the session therefore took place in a standard museum workroom, characterized by arbitrary acoustics generally unsuitable for recording. Available playing time was strictly limited: aged wood is highly sensitive to humidity, and sustained playing introduces moisture at the embouchure, creating a real risk of structural damage. This ruled out thorough coverage of dynamic range, repeated takes, and a variety of extensive articulations typical in sample library production. The session was organized to maximize acoustically informative yield within a minimal time window, ultimately lasting only a few minutes.

The recording focused on capturing sustained timbres and attack transients (tone-onsets), performed by traverso expert performer Kate Clark. Attack transients carry important perceptual information for instrument identification~\cite{cassidy2016role}. In historical traversos they are dominated by the \textit{chiff}, a brief inharmonic noise burst from turbulent airflow at the embouchure edge before the fundamental is established~\cite{castellengo1999acoustical}. 
While the chiff is present in all flutes, it is especially pronounced in thick-walled wooden instruments, where the thick embouchure edge softens the division of the airstream compared to the thin lip plate of a modern metal flute.
The notes were recorded chromatically at a single \textit{mezzo-forte} dynamic, across A\#3 (233~Hz) -- B4 (486~Hz) per standard tuning, which corresponds to C\#4--D5 at the Haka's tuning of A\,=\,365~Hz. Supplemental recordings were conducted on a modern Haka replica, as a second recording session with the original instrument was logistically impractical.

\subsection{Spectral Analysis}

\begin{figure}[htbp]
  \centering
  \begin{subfigure}[b]{0.48\columnwidth}
    \centering
    \includegraphics[width=\textwidth]{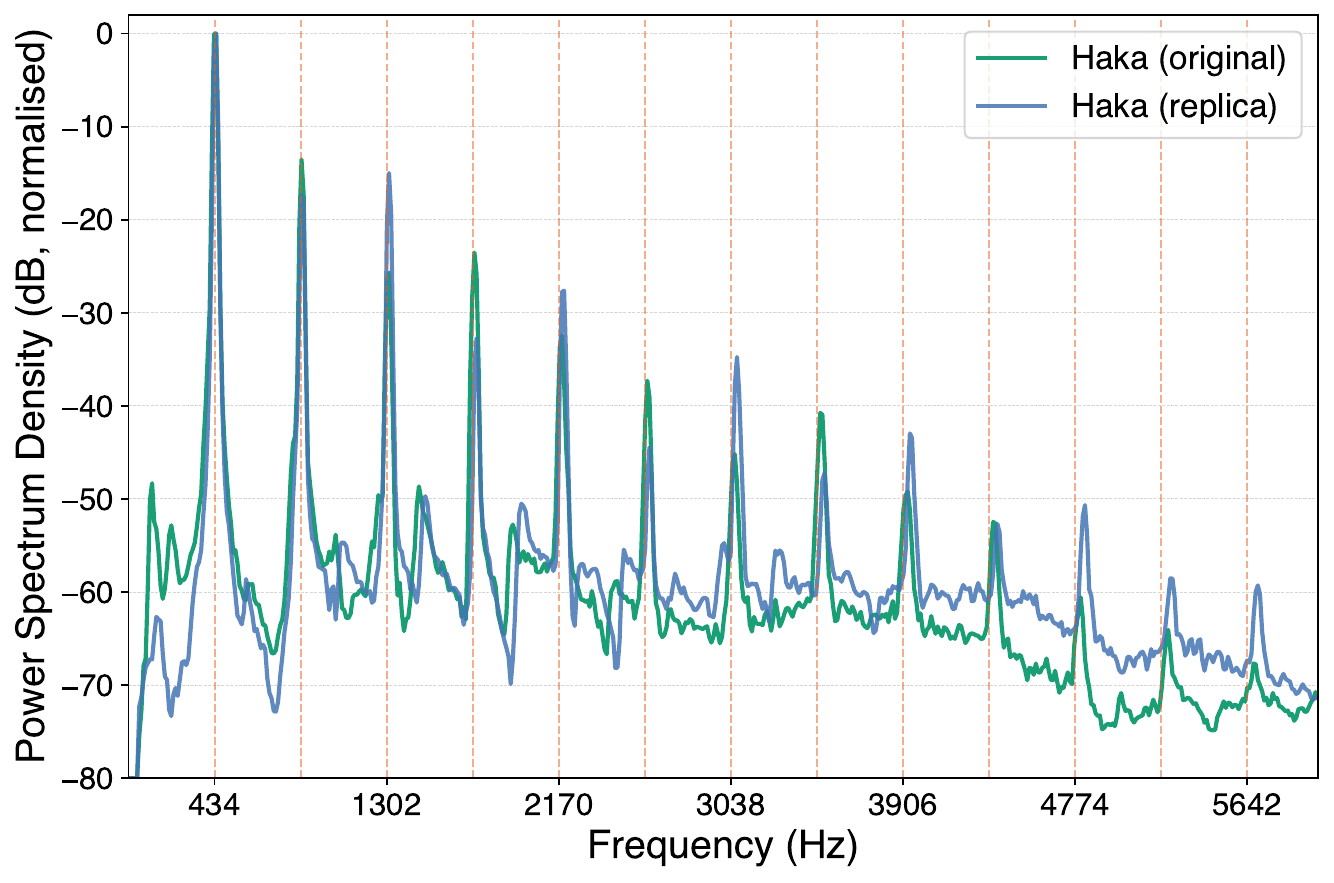}
    \caption{The Haka flute}
    \label{fig:haka_spectra_a}
  \end{subfigure}
  \hfill
  \begin{subfigure}[b]{0.48\columnwidth}
    \centering
    \includegraphics[width=\textwidth]{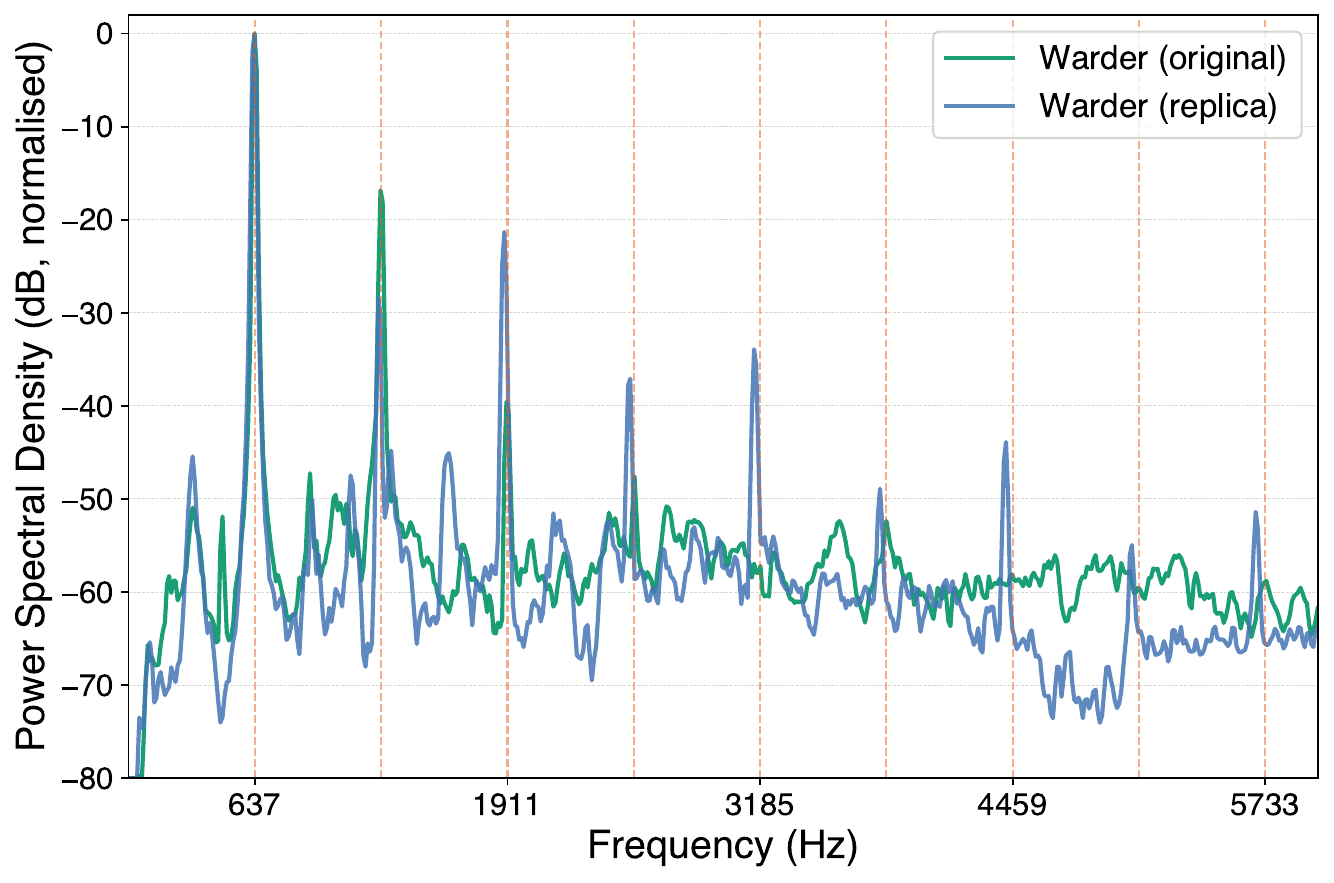}
    \caption{The Warder flute}
    \label{fig:haka_spectra_b}
  \end{subfigure}
  \caption{Power spectral density (Welch's method) of original instrument vs.\ replica. (a)~Haka flute, note C5 (Haka's pitch): the original shows greater power at the 2nd, 4th, 6th, and 8th harmonics; content drops near the noise floor beyond the 10th harmonic. (b)~Warder flute, note F\#5 (Warder's pitch): the original shows greater power at the 2nd harmonic, while the replica is stronger at the 3rd and 4th; the original drops near the noise floor beyond the 4th harmonic.}
  \label{fig:haka_spectra}
\end{figure}

A comparison of the Welch's method power spectral density of the original Haka flute and its modern replica, for the note C5 (434~Hz, at Haka's pitch) is shown in
Fig.~\ref{fig:haka_spectra_a}. Both tones show a comparable power at the fundamental frequency, with the original showing larger power at the 2nd, 4th, 6th and 8th harmonics. This behaviour is also observed in most additional Haka tones. Beyond approximately the 10th harmonic, the original flute's power diminished to near the noise floor, most likely due to recording conditions.

\subsection{Digital Instrument Design}
\label{sec:design}

The samples acquired from the Haka flute were used to design a playable digital instrument. It combines three layers per note: (1) sustained tones sampled from the original Haka flute; (2) a physical model realized in Respiro, an audio plugin for woodwind physical modelling; and (3) a dedicated attack tone layer. The instrument spans five octaves (C1--C6 in Haka's pitch), extended by pitch-transposition. The instrument was implemented in Kontakt with the explicit intention of being performed via an EWI.

The three-layer architecture reflects both technical constraints and the aesthetic choices of the performing musician, and is affected by both the EWI's mode of operation and the nature of sample libraries. The EWI operates in legato mode, in which a new note onset overlaps briefly with the tail of the preceding note. In sample-based instruments this overlap is typically handled by a crossfade script, which also suppresses the attack transient. The dedicated attack layer resolves this issue, preserving attack identity even under legato playing configuration. The Respiro layer fills two expressive gaps that sample playback cannot cover: At very low breath pressure, samples fall below their response threshold and produce silence, whereas a real flute responds subtly from the first movement of air. At very high breath pressure, a physical flute overblows and produces higher harmonics, which do not appear in the samples.

The unique flute characteristics are activated via the EWI's breath control, which is mapped simultaneously to the output volume and to a low-pass filter cutoff frequency. Thus, increasing breath pressure raises both amplitude and spectral brightness, approximating the live behavior of a wind instrument.

\subsection{Artistic Outcome}

The revived digital instrument is featured in the album \ifanonymous [anonymized for review]\else ``Soundfront 1: The Richard Haka Flute'' by Itai Weissman (Conservatorium van Amsterdam)\fi~\ifanonymous \cite{anon2025album} 
\else \cite{weissman2025soundfont} 
\fi. The album features the revived digital instrument performed via EWI, as well as a physical Haka replica flute, a harp, and spoken narration.

The musical material was developed through extended improvisation sessions between the EWI and harp, subsequently edited into finished pieces with additional layered tracks. The compositional approach deliberately departs from historically informed performance practice. It explores possibilities the original instrument never offered: an expanded range, polyphony, and a freely improvisational idiom foreign to the flute's own era. The album can thus be understood as a reflection on what a historical sound source might become when removed from its original context and placed in the hands of contemporary creative practice.

\section{Digital Revival of a Non-Functional Flute: Warder Case Study}

This case study presents an ongoing digital revival project centered on the Warder flute, a Renaissance transverse flute that is almost entirely unplayable due to structural damage. In contrast to the Haka case, where a functional instrument permitted direct acoustic recording, the Warder project must rely primarily on physical measurements, scan data, physical modelling, and modern replicas.

\subsection{The Warder Flute}

The Warder flute is a tenor Renaissance traverso dated to the 1540s, recovered in a waterlogged condition in 2017 from a shipwreck off the coast of Warder, North Holland~\cite{huisman2020warder}, 
and is considered the oldest surviving flute found in the Netherlands. It is now held at Huis van Hilde, the Archaeological Museum of North Holland in Castricum.

The instrument measures 69~cm in length, has six undercut finger holes, and is carved from a single piece of boxwood. It bears visible signs of past repair: seven bands of iron wire near the mouthpiece mend a crack, and a brass tube at the mouthpiece end contains a fragment of printed paper in German Gothic typeface. The paper is likely from a 16th-century Nuremberg music publication, suggesting the instrument may have belonged to a mercenary in West-Friesland~\cite{huisman2020warder}.

\begin{figure}[htbp]
  \centering
  \begin{subfigure}[b]{0.66\columnwidth}
    \centering
    \begin{subfigure}[b]{0.31\linewidth}
      \includegraphics[width=\linewidth, trim={0 14pt 0 14pt}, clip]{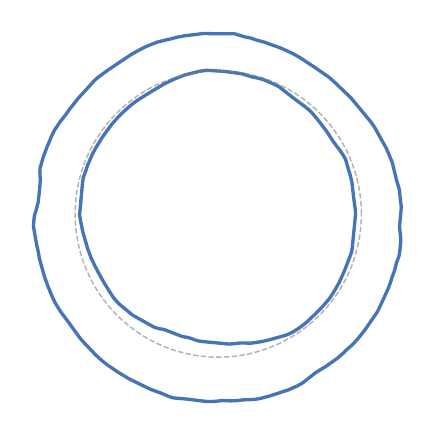}
      \caption*{\footnotesize 40 mm}
    \end{subfigure}\hfill
    \begin{subfigure}[b]{0.31\linewidth}
      \includegraphics[width=\linewidth, trim={0 14pt 0 14pt}, clip]{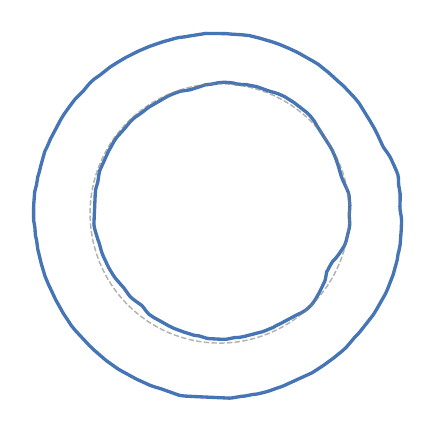}
      \caption*{\footnotesize 160 mm}
    \end{subfigure}\hfill
    \begin{subfigure}[b]{0.31\linewidth}
      \includegraphics[width=\linewidth, trim={0 14pt 0 14pt}, clip]{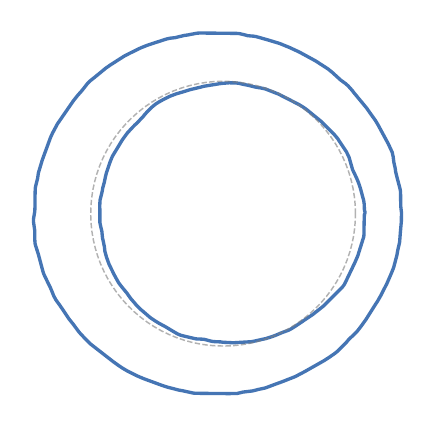}
      \caption*{\footnotesize 280 mm}
    \end{subfigure}

    \vspace{3pt}

    \begin{subfigure}[b]{0.31\linewidth}
      \includegraphics[width=\linewidth, trim={0 14pt 0 14pt}, clip]{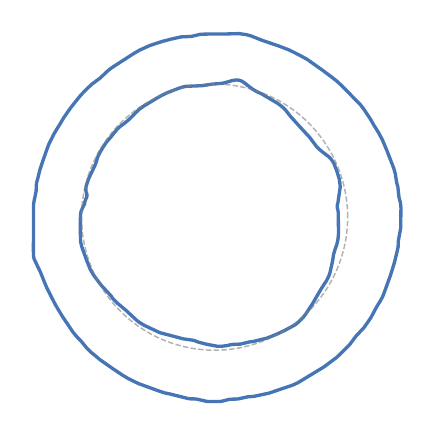}
      \caption*{\footnotesize 400 mm}
    \end{subfigure}\hfill
    \begin{subfigure}[b]{0.31\linewidth}
      \includegraphics[width=\linewidth, trim={0 14pt 0 14pt}, clip]{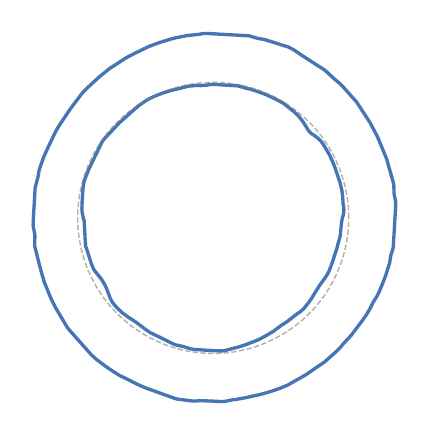}
      \caption*{\footnotesize 520 mm}
    \end{subfigure}\hfill
    \begin{subfigure}[b]{0.31\linewidth}
      \includegraphics[width=\linewidth, trim={0 14pt 0 14pt}, clip]{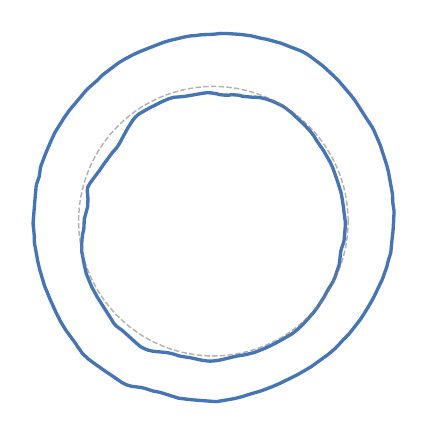}
      \caption*{\footnotesize 640 mm}
    \end{subfigure}
    \caption{}
    \label{fig:cross_sections_a}
  \end{subfigure}%
  \hfill
  \begin{subfigure}[b]{0.30\columnwidth}
    \centering
    \includegraphics[width=0.95\linewidth, trim={0 100pt 0 950pt}, clip]{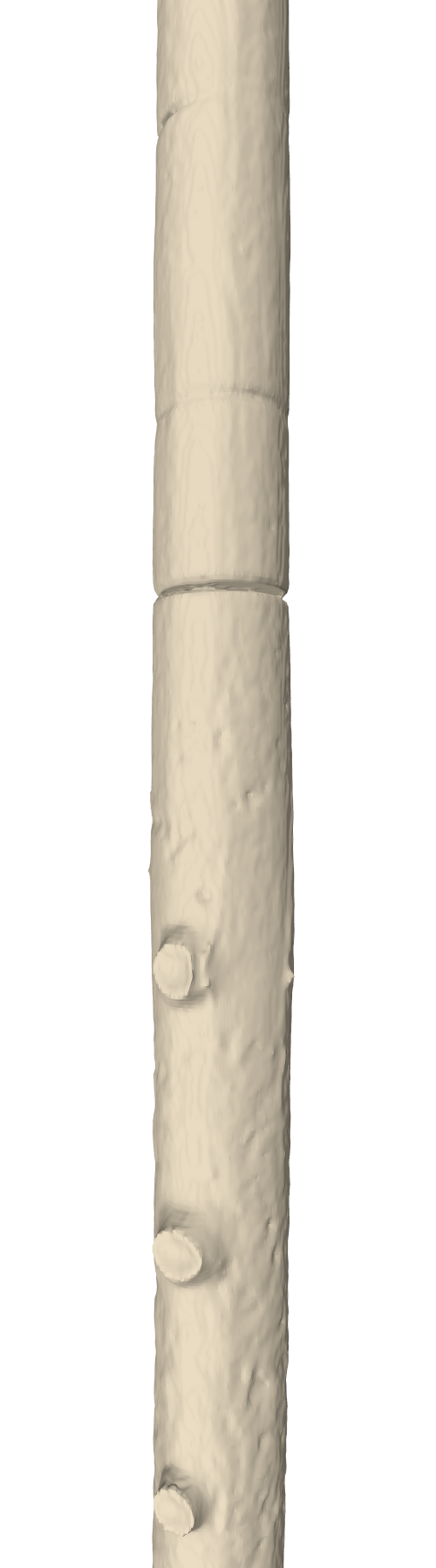}
    \caption{}
    \label{fig:cross_sections_b}
  \end{subfigure}
  \caption{(\subref{fig:cross_sections_a})~Cross-sections of the Warder flute wall at six positions along the bore axis, derived from the 3D scan. Distances are measured from the open end. The dashed circle (20~mm diameter) serves as a reference. (\subref{fig:cross_sections_b})~3D render of the inner bore surface at the topmost tone holes, showing the irregular and deformed surface.}
  \label{fig:cross_sections}
\end{figure}

Five centuries of submersion, followed by controlled post-excavation drying, have substantially altered the wood's mechanical properties. CT scans reveal that the bore cross-section is oval and irregular, consistent with waterlogged burial and post-excavation drying, with direct acoustic consequences. The bore axis also exhibits a subtle longitudinal curvature. Detailed measurement and CT scanning were carried out by researchers at TU Delft~\cite{huisman2020warder}, after which flute-builder Roberto Bando~\cite{bando2025flutes} constructed two replicas: a first in maple to verify dimensions, and a second in boxwood. Warder's cross-section plots and 3D model of the internal cavity are shown in Fig.~\ref{fig:cross_sections}.

\subsection{Recording Procedure and Challenges}

The flute's current structural state severely limits playability. Damage includes multiple longitudinal cracks, localized bore deformation, and mechanically weakened wood fibers. As a result, only two notes could be produced during the recording session: F\#5 (637~Hz) and G5 (685~Hz), at the Warder's pitch of A\,=\,386~Hz. The session was conducted in the same room as the Haka session and similarly constrained in duration. Non-invasive acoustic measurements such as input impedance were not feasible, as the instrument's fragility and archive access restrictions make transport to a laboratory setting impractical.

\subsection{Available Data and Acoustic Characterization}

Given Warder's condition, acoustic characterization draws on two complementary sources: the limited direct recordings and a comprehensive set of physical measurements.

The physical dataset includes 3D models of the instrument exterior and bore interior, volumetric IMA scan files containing CT-like cross-sectional images of the bore, and a spreadsheet of quantitative geometric measurements such as bore diameters, cross-sectional min/max dimensions, and wall thickness. Together these characterize the bore geometry, wall profile, and axial curvature along the full length of the instrument. Two replicas have been constructed from this data and are available for extended sampling.

\subsection{Spectral Analysis}

A comparison of the power spectral density, estimated using Welch's method, of the Warder flute and its replica, for the note F\#5 (637~Hz, at Warder's pitch), is shown in Fig.~\ref{fig:haka_spectra}. The fundamental frequency is comparable between the two instruments. The Warder shows higher power at the 2nd harmonic, while the 3rd and 4th harmonics are stronger in the replica. Beyond the 4th harmonic, both signals drop to near the noise floor, most likely due to recording conditions.

\subsection{Digital Revival Approach}

The digital revival of the Warder flute proceeds along two parallel tracks. The first is replica-based sampling: since the replicas are fully playable, they can be sampled using Haka's methodology, providing a complete note inventory and articulation set, which will be used either as a reference, or integrated into a future digital instrument. The second track is geometry-based physical modeling: the scan data provides sufficient geometric information to parameterize a physical model of the air column, from which the acoustic behavior of the original instrument can be estimated. The digital instrument design will most likely follow the same layered architecture as the Haka case, adapted to the different dataset. The Warder revival is currently in early stages, with replica sampling underway and physical modelling in preparation.

\section{Discussion}

This paper has presented two case studies in the Digital Revival framework. The Haka flute is a functional museum instrument sampled under conservation constraints and reactivated as a playable digital instrument; the Warder flute is a structurally compromised archaeological find whose revival relies on scan data, physical modelling, and modern replicas. Both cases illustrate the challenges of acoustic revival.

\subsection{Authenticity vs.\ Expressivity: The Haka Case}

A central tension in the Haka instrument design is the trade-off between acoustic authenticity and expressive capability. Sample-based playback can faithfully reproduce the timbral character of recorded tones, but it cannot capture the full dynamic behavior of a wind instrument. At very low breath pressure, samples fall below their response threshold. At high pressure, a real flute overblows into upper harmonics. These behaviors are absent from a purely sample-based instrument. The physical model layer (Respiro) was included precisely to bridge this gap, providing a continuous response across the full breath range.

The recording conditions further complicated the pursuit of authenticity. An extremely short session in a non-dedicated space yielded recordings lacking full high frequency content, which needed to be compensated by using an additional layer of samples obtained from the replica. The resulting digital instrument, with its three layers, therefore reflects a balance among three considerations: the acoustic character of the original instrument, the expressive requirements of real-time EWI performance, and the aesthetic judgment of the performing musician. This trade-off is not a failure of the methodology but an inherent feature of a project that operates under real-world constraints while also pursuing an artistic, musical goal.

\subsection{What Can Be Recovered: The Warder Case}

The Warder flute poses a deeper epistemological challenge: it is unknown, and cannot be directly determined, what the instrument sounded like originally. Five centuries of submersion, burial, and post-excavation drying have altered virtually every acoustically relevant property of the object at least to some extent.

The Warder's deformed bore cross-section is the primary acoustic concern. The original bore profile is unknown and was possibly imperfect even when new, given the fabrication techniques of the period; the extent of the deformation attributable to burial cannot therefore be determined. The flute is also slightly curved along its axis, but this is expected to have negligible acoustic consequence, and there is no reason to believe it was a feature of the original instrument.

The condition of the tone holes presents a further interpretive difficulty. The sharpness of the embouchure edge and the geometry of the finger holes are acoustically critical, yet their original form cannot be determined. A blunt or deformed edge observed today may have been originally sharp and precise. Since the maker is unknown, there is no workshop tradition or comparative corpus to draw on.

This raises a fundamental artistic question that the Digital Revival project must answer explicitly: \textit{is the instrument to be reconstructed as it was originally, or as it is now?}

\subsection{The Material Unknown}

Although bore geometry is the primary acoustic determinant in flutes, material properties are somewhat relevant: the material affects the instrument's tactile response and thus indirectly the playing, and may also have some direct acoustic effect. The instruments cannot be subjected to mechanical testing or sampled for chemical analysis. While non-invasive analysis methods may be suitable, they require a laboratory setting, which is logistically impractical. Material identification is therefore limited to visual inspection. The material was identified as boxwood, but the acoustic properties of the specific piece, such as density and damping characteristics, can only be estimated. Wood aging further complicates this: centuries-old boxwood resonates differently from freshly cut material, and waterlogged wood differs again. The revival therefore proceeds on the assumption that geometry dominates, while acknowledging that material uncertainty places a fundamental limit on acoustic fidelity that no amount of geometric precision can resolve.

\section*{Acknowledgements}
The authors wish to thank Giovanni Paolo Di Stefano (Rijksmuseum, instruments curator), Professor Alon Schab (Bar-Ilan University, external advisor), Kate Clark (Royal Conservatoire of The Hague, historical flutes), Caro Verbeek (Kunstmuseum Den Haag, curator), Johan Looijenga (Utrecht University, Lecturer, Industrial Engineering), and Michiel Schuijer (Conservatorium van Amsterdam, external advisor) for their advice and support throughout this research.

\newpage
\bibliographystyle{unsrt}
\bibliography{references}

\end{document}